\documentclass[sigconf,natbib=true]{acmart}
\usepackage{nccmath} 
\usepackage{algorithm}
\usepackage{algorithmic}
\usepackage{color}
\usepackage{graphicx}
\graphicspath{{images/}}
\usepackage{wrapfig,lipsum}
\usepackage{tabularx}
\usepackage{graphicx}
\usepackage{lscape}
\usepackage{subcaption}
\usepackage{latexsym}
\usepackage{pifont}
\usepackage{multirow}

\newcommand{\cmark}{\ding{51}}%
\usepackage{enumitem}
\setlist{leftmargin=3mm}
\usepackage[T1]{fontenc}
\usepackage[utf8]{inputenc}
\usepackage[font=small,labelfont=bf]{caption}

\usepackage{booktabs}

\usepackage{xcolor}
\usepackage{xspace}
\usepackage{bm}
\usepackage{amsmath}

\AtBeginDocument{%
  \providecommand\BibTeX{{%
    \normalfont B\kern-0.5em{\scshape i\kern-0.25em b}\kern-0.8em\TeX}}}

\copyrightyear{2023}
\acmYear{2023}
\setcopyright{licensedothergov}\acmConference[SIGIR-AP '23]{Annual International ACM SIGIR Conference on Research and Development in Information Retrieval in the Asia Pacific Region}{November 26--28, 2023}{Beijing, China}
\acmBooktitle{Annual International ACM SIGIR Conference on Research and Development in Information Retrieval in the Asia Pacific Region (SIGIR-AP '23), November 26--28, 2023, Beijing, China}
\acmPrice{15.00}
\acmDOI{10.1145/3624918.3625324}
\acmISBN{979-8-4007-0408-6/23/11}

\begin{document}
\title[Typos-aware Bottlenecked Pre-Training for Robust Dense Retrieval]{Typos-aware Bottlenecked Pre-Training\\ for Robust Dense Retrieval}


\settopmatter{printacmref=true, authorsperrow=4}

\author{Shengyao Zhuang}
\affiliation{%
	\institution{\mbox{The University of Queensland}}
	\city{Brisbane}
	\country{Australia}}
\email{s.zhuang@uq.edu.au}
\authornote{Work done during internship at Microsoft STCA.}

\author{Linjun Shou}
\affiliation{
	\institution{Microsoft}
	\city{Beijing}
	\country{China}}
\email{lisho@microsoft.com}

\author{Jian Pei}
\affiliation{%
	\institution{Duke University}
	\city{Durham}
	\country{USA}}
\email{j.pei@duke.edu}

\author{Ming Gong}
\affiliation{
	\institution{Microsoft}
		\city{Beijing}
	\country{China}}
\email{migon@microsoft.com}

\author{Houxing Ren}
\affiliation{
	\institution{Beihang University}
		\city{Beijing}
	\country{China}}
\email{renhouxing@buaa.edu.cn}
\authornotemark[1]

\author{Guido Zuccon}
\affiliation{
	\institution{\mbox{The University of Queensland}}
	\city{Brisbane}
	\country{Australia}}
\email{g.zuccon@uq.edu.au}
\authornote{Corresponding authors.}

\author{Daxin Jiang}
\affiliation{
	\institution{Microsoft}
		\city{Beijing}
	\country{China}}
\email{djiang@microsoft.com}
\authornotemark[2]


\begin{abstract}

Current dense retrievers (DRs) are limited in their ability to effectively process misspelled queries, which constitute a significant portion of query traffic in commercial search engines. 
The main issue is that the pre-trained language model-based encoders used by DRs are typically trained and fine-tuned using clean, well-curated text data. Misspelled queries are typically not found in the data used for training these models, and thus misspelled queries observed at inference time are out-of-distribution compared to the data used for training and fine-tuning. Previous efforts to address this issue have focused on \textit{fine-tuning} strategies, but their effectiveness on misspelled queries remains lower than that of pipelines that employ separate state-of-the-art spell-checking components.

To address this challenge, we propose ToRoDer (TypOs-aware bottlenecked pre-training for RObust DEnse Retrieval), a novel \textit{pre-training} strategy for DRs that increases their robustness to misspelled queries while preserving their effectiveness in downstream retrieval tasks. ToRoDer utilizes an encoder-decoder architecture where the encoder takes misspelled text with masked tokens as input and outputs bottlenecked information to the decoder. The decoder then takes as input the bottlenecked embeddings, along with token embeddings of the original text with the misspelled tokens masked out. The pre-training task is to recover the masked tokens for both the encoder and decoder.

Our extensive experimental results and detailed ablation studies show that DRs pre-trained with ToRoDer exhibit significantly higher effectiveness on misspelled queries, sensibly closing the gap with pipelines that use a separate, complex spell-checker component, while retaining their effectiveness on correctly spelled queries. 

\end{abstract}


\maketitle

\section{Introduction}

Dense Retrievers (DRs) have been shown to have higher effectiveness than traditional keyword-based matching methods such as BM25 on information retrieval (IR) tasks~\cite{tonellotto2022lecture,zhao2022dense}. Aside from the fine-tuning process that provides a mean for these representations to be learned on the task and dataset at hand, the effectiveness gains provided by DRs can be largely attributed to the semantics encoded in the underlying large pre-trained language models these DRs are built from. There is mounting evidence that DRs are currently being integrated in many commercial search solutions because of their demonstrated effectiveness~\cite{kim2022applications,du2022open,benedetti2022dense,huang2020embedding}.

Although highly effective for in-domain data -- that is, test data that is drawn from the same distribution as the training data --  previous studies have reported that DRs are less robust to out-of-domain data.
For example, DRs trained on the MS MARCO dataset (web data) are less effective than BM25 on the TREC-COVID collection (medical literature data)~\cite{chen2022out,thakur2021beir}.

In this paper, we focus on a challenging special case of out-of-domain data for DRs: misspelled queries. This is an important issue as misspellings are frequently encountered in users queries~\cite{nordlie1999user,spink2001searching,wang2003mining,wilbur2006spelling,hagen2017large}. Previous work has shown that the effectiveness of DRs is dramatically reduced when typos are present in the search queries~\cite{zhuang2021dealing,zhuang2022char,sidiropoulos2022analysing,Chen2022towards,penha2022evaluating,wu2022neural,zhuang2022robustness,penha2022evaluating}. 
We argue that this is mainly because: (1) the DRs encoders are based on pre-trained language models for which the pre-training occurred on curated text with no or little misspellings (e.g., Wikipedia and books); and (2) current DRs are also fine-tuned with curated datasets -- for example, MS MARCO has no or little misspelled queries in its training data. Thus, in these cases, test queries that contain misspellings become  out-of-distribution with respect to the data used for training.

Current approaches that have attempted to improve DRs effectiveness on misspelled queries do so in the fine-tuning stage~\cite{zhuang2021dealing,zhuang2022char,sidiropoulos2022analysing,Chen2022towards} -- thus addressing point 2 above. These methods are generally referred to as \textit{typos-aware fine-tuning}.
 For instance, \citeauthor{zhuang2021dealing}~\cite{zhuang2021dealing} were the first to investigate this problem; they proposed a data augmentation-based DR fine-tuning strategy which randomly injects typos into training queries. Training queries, now containing a mix of queries with and without typos, are then used within a standard DR fine-tuning pipeline. Later work then has built upon this approach by proposing novel loss functions to align the embeddings of queries without typos with  the embeddings of the corresponding misspelled queries~\cite{zhuang2022char,sidiropoulos2022analysing,Chen2022towards}.
Although these approaches significantly improve the robustness of the DRs on misspelled queries, their effectiveness on misspelled queries is still much lower than if the typo was not present in that same query.


No current approach has considered addressing the point 1 above: the fact that the pre-training of the language models used to create DRs occurs on data that seldom contains typos. But, how to design a pre-training strategy that can make DRs effective for the downstream retrieval task while remaining robust to misspelled queries? This is what we set to address in this paper. To this end, we develop a pre-training strategy purposely designed for dense retrievers to handle misspelled queries. This strategy can be further coupled with typo-aware fine-tuning strategies.
Our strategy, \textit{ToRoDer}: TypOs-aware bottlenecked pre-training for Robust DEnse Retrieval, adapts an encoder-decoder architecture where the encoder is a full-size BERT model (12 transformer layers) and the decoder is a smaller and weaker BERT model (2 transformer layers). The strategy builds upon recent advances in dense bottleneck-enhanced pre-training~\cite{gao-callan-2021-condenser,gao2022unsupervised,lu-etal-2021-less,liu2022retromae,wang2022simlm,shen2022lexmae,chuang2022diffcse,wu2022contextual}.
During pre-training, the encoder takes misspelled text with some tokens masked out as inputs and outputs [CLS] embeddings as bottlenecked information to the decoder. The decoder then takes the encoded embeddings as input along with the token embeddings of the original text.  Compared to the encoder input, the text given as input to the decoder has the same [MASK] tokens as the encoder input but misspelled tokens are now also replaced with the [MASK] tokens. The encoder-decoder is then trained end-to-end with the \textit{Masked Language Modelling} (MLM) approach, where the decoder needs to predict the masked-out tokens as well as the original, correctly spelled tokens corresponding to the misspelled tokens. 
Since the decoder is just a two-layer small transformer model, it has very limited modelling power. Therefore, in order to help the decoder recover the original, correctly spelled tokens, the encoder needs to compress as much information as possible into the bottlenecked [CLS] token embeddings. 

We carry out an extensive suite of experiments on both publicly available benchmark datasets and our internal dataset from a large commercial search engine. The empirical results show that the DR pre-trained with our proposed ToRoDer and fine-tuned with standard hard negative sampling, knowledge distillation, and typos-aware fine-tuning can achieve the same level of effectiveness of state-of-the-art dense retrievers; however, at the same time, it also demonstrates much higher effectiveness on misspelled queries.
Our ablation analysis further suggests that our typos-aware bottlenecked pre-training is crucial for DRs to achieve the highest effectiveness on misspelled queries. 

Our main contributions are:
\begin{enumerate}
	\item We are the first to propose a pre-training approach to enhance the robustness of dense retrievers, building upon the existing bottleneck pre-training technique. This contribution is significant as it addresses a key challenge in dense retrieval systems.
	\item We introduce a unique integration of typo generation with the MLM task during bottlenecked pre-training for dense retrievers. Our study includes a comprehensive investigation of the impact of typo generation rate on retrieval effectiveness, providing valuable insights for practitioners in the field.
	\item Our extensive experimental results on both synthetic and real-world datasets demonstrate the superior effectiveness of our approach, with the key finding that using synthetic training data is effective to answer real queries with typos.
\end{enumerate}

\vspace{-8pt}
\section{Related works}
\subsection{Effectiveness and robustness of DRs}
Traditional keyword matching or bag-of-words methods for retrieval, such as BM25, are affected by the vocabulary mismatch problem: relevant passages that use semantically equivalent but different words are typically not retrieved~\cite{furnas1987vocabulary}. Dense retrievers (DRs)~\cite{zhao2022dense} leverage pre-trained language models  such as BERT~\cite{DBLP:conf/naacl/DevlinCLT19}, RoBERTa~\cite{liu2019roberta} and others, to embed both queries and documents into a dense vector space, and then compute their relevance score by some distance metric such as dot product. These dense retrievers have been shown to be much more effective than exact keyword matching based methods when enough training data is provided~\cite{karpukhin-etal-2020-dense,xiong2020approximate,zhan2020repbert,Zhan2021OptimizingDR,ren2021pair,ren2021rocketqav2,qu2021rocketqa,lin2021batch,lin2020distilling,hofstatter2020improving,hofstatter2021efficiently}. However, recent studies have shown that DRs may be less effective under certain circumstances. For instance, \citet{mackie2021how} introduced a framework for identifying hard queries that pose a challenge for DRs, and \citet{sciavolino2021simple} found that DRs perform poorly, and worse than traditional keyword matching methods, on queries that contain entities. Additionally, results on the BEIR zero-shot IR benchmark dataset indicate that most DRs perform worse than BM25 on out-of-domain datasets~\cite{chen2022out,thakur2021beir}. An in-depth examination of these results has revealed that DRs excel when the overlap between queries and passages is low, however, they exhibit considerable losses compared to BM25 when the overlap is large~\cite{ren2022thorough}.

\vspace{-10pt}
\subsection{Robustness of DRs on misspelled queries}
A particular aspect related to the robustness of DRs is their effectiveness on the corresponding correctly spelled queries~\cite{zhuang2021dealing,zhuang2022char,sidiropoulos2022analysing,Chen2022towards,penha2022evaluating,wu2022neural,zhuang2022robustness,penha2022evaluating} -- this setting is the focus of our paper. 

\citet{zhuang2021dealing} first showed that the effectiveness of DRs trained with standard contrastive learning degrades significantly when answering misspelled queries. This problem was found to also extend to PLM-based sparse retrieval methods~\cite{zhuang2022robustness}. In addition to typos, various other query variations were shown to have a negative impact on the effectiveness of DRs~\cite{Chen2022towards}. To address the lack of robustness on misspelled queries, \citet{zhuang2021dealing} proposed a data augmentation-based method that uses synthetic generated queries in the fine-tuning of the rankers. \citet{sidiropoulos2022analysing} further suggested that combining data augmentation with contrastive learning can generate representations of a query with typos that are close to those of the corresponding queries without typos, thus improving DR effectiveness on the misspelled queries. At the same time, this observation was made also by \citet{zhuang2022char} who identified that the tokenizer of BERT is sensitive to typos and CharacterBERT \cite{el2020characterbert} is a better backbone PLM model for training a typo-robust DR. In the same work, they also proposed a novel typos-aware fine-tuning approach called Self-Teaching (ST) which distillates the score distributions of the DR on correctly spelled queries to its misspelled queries. The combination of CharacterBERT and ST constitutes the current state-of-the-art in DRs robust to queries with typos. A similar method was later devised by \citet{Chen2022towards}, who introduced the RoDR model which maintains the relative positions of query-passage pairs for original and misspelled queries; for this, it uses a loss similar to ST. It is noteworthy that all the aforementioned methods focus on addressing the lack of robustness to typos in the DR's fine-tuning stage, while ToRoDer focuses on the pre-training stage.

\vspace{-5pt}
\subsection{Bottlenecked pre-training for DRs}

Most pre-training tasks for PLMs are designed without any
prior knowledge of the downstream dense retrieval task: thus general PLMs such as BERT may not be optimal for the dense retrieval task. To address this and to further improve the effectiveness of PLM-based DRs, recent studies have explored bottlenecked pre-training approaches~\cite{gao-callan-2021-condenser,gao2022unsupervised,lu-etal-2021-less,liu2022retromae,wang2022simlm,shen2022lexmae,chuang2022diffcse,wu2022contextual}. These approaches leverage encoder-decoder architectures to produce better [CLS] token representations within the PLMs used for dense retrieval. For example, the Condenser architecture proposed by~\citet{gao-callan-2021-condenser} splits the PLM into early and late layers, with a weaker decoder trained to refine the [CLS] token representation from the late layer of the PLM. Similarly, SEED~\cite{lu-etal-2021-less} employs an encoder-decoder architecture with the [CLS] token trained using an autoregressive next token prediction objective. More recently,  SimLM~\cite{wang2022simlm} utilizes an ELECTRA-style replaced language modeling objective~\cite{clark2020electra} which uses a generator to sample replaced tokens for
	masked positions, and then uses a bottlenecked encoder-decoder architecture to predict the original tokens in
	all positions. In this work, we adapt a similar bottlenecked pre-training approach to enhance the robustness of the [CLS] token representation to misspelled texts.

\begin{figure}[t]
\centering
\includegraphics[width=0.75\columnwidth]{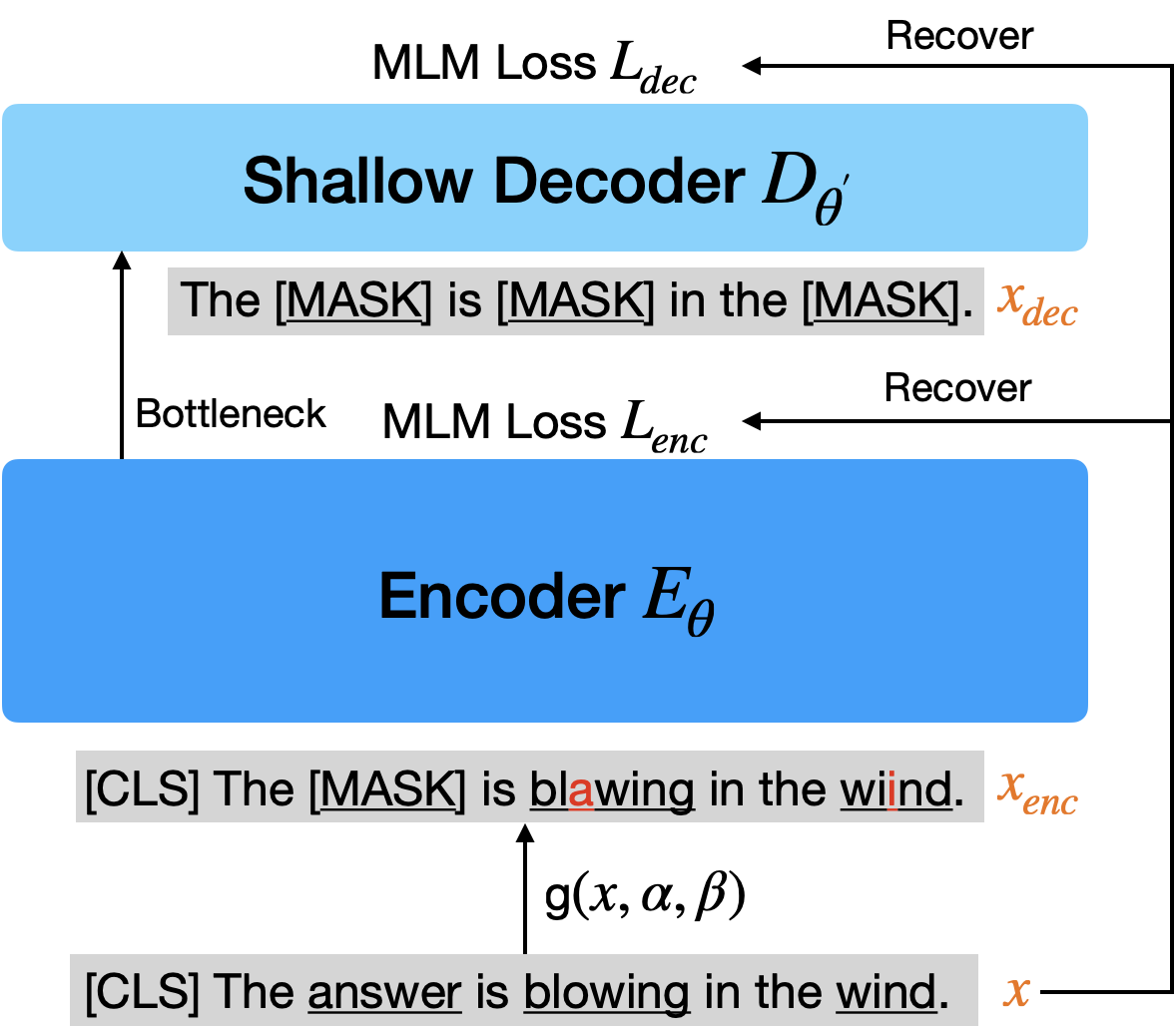}
\vspace{-10pt}
\caption{The architecture of our proposed typos-aware bottlenecked pre-training approach, ToRoDer. The input text $x$ is modified by the text editor $g$, which is controlled by parameters $\alpha$ (masking ratio) and $\beta$ (typo ratio), to produce the encoder input $x_{enc}$. The decoder input  $x_{dec}$ consists of an encoder outputted bottlenecked [CLS] embedding plus the text $x$ with all the modified tokens replaced with [MASK]. The losses $\mathcal{L}_{enc}$ and $\mathcal{L}_{dec}$ aim to recover the original tokens in $x$ in place of the [MASK] tokens. }
\label{fig:1}
\vspace{-12pt}
\end{figure}
\vspace{-4pt}
\section{Typos-aware bottlenecked pre-training}\label{sec:pretrain}

Previous works have only focused on the fine-tuning stage to improve the effectiveness and robustness of DRs on misspelled queries~\cite{zhuang2021dealing,zhuang2022char,sidiropoulos2022analysing,Chen2022towards}. We believe the pre-training stage is also important for training a typo-robust DR. This is intuitive because the self-supervised pre-training tasks for the backbone language models have a significant influence on the downstream tasks (in our case, the passage retrieval task)~\cite{clark2020electra,DBLP:conf/naacl/DevlinCLT19,liu2019roberta}. Hence, our goal is to design a pre-training task for the DRs' backbone model that not only benefits the downstream retrieval task, but is also tolerant to misspelled text.
Inspired by the recent advances in dense bottleneck-enhanced pre-training~\cite{gao-callan-2021-condenser,gao2022unsupervised,lu-etal-2021-less,liu2022retromae,wang2022simlm,shen2022lexmae,chuang2022diffcse,wu2022contextual}, which bring significant improvements to the effectiveness of DRs, we propose ToRoDer: Typos-aware bottlenecked pre-training for robust dense retrieval. 

Figure~\ref{fig:1} illustrates our ToRoDer pre-training approach. It has two components, a text encoder $E_\theta$ and a shallow decoder $D_{\theta'}$. 
We instantiate the encoder with a deep multi-layer Transformer model~\cite{vaswani2017attention} that can be initialized with pre-trained models such as BERT~\cite{DBLP:conf/naacl/DevlinCLT19}; this is in line with most previous works. The decoder has the same architecture as the encoder but with just two Transformer layers (as opposed to the 12 layers of the encoder in its BERT initialization). In addition, we initialize the decoder parameters $\theta'$ with the word embeddings and the last two Transformer layers of the encoder.

During pre-training, given an input text $x$, we use a rule-based ``text editor'' $g$ to modify $x$ and obtain the encoder text input $x_{enc}$. The text editor does two types of modifications to the original text: 
\begin{enumerate}
	\item the masking of sampled tokens based on the masked language modeling (MLM) approach, in line with its use in BERT~\cite{DBLP:conf/naacl/DevlinCLT19}, and
	\item the replacement of sampled tokens with a modified token in which a typo has been injected according to the typo generation methods used by Zhuang and Zuccon~\cite{zhuang2021dealing}.
\end{enumerate}

 
Specifically, for the MLM task, $\alpha\%$ of the tokens in $x$ are sampled and 80\% of the times they are replaced with the special token [MASK], 10\% of the times they are replaced with a random token in the vocabulary, and the remaining 10\% are kept unchanged~\cite{DBLP:conf/naacl/DevlinCLT19}. 
For the typo generation task, $\beta\%$ of the sampled tokens are modified by injecting a typo. A typo is injected by randomly sampling from five typo types (RandInsert, RandDelete, RandSub, SwapNeighbor, and SwapAdjacent) in accordance to prior work~\cite{zhuang2021dealing}. We name the parameter $\beta$ as the typo injection ratio.
Similar to the MLM task, the typo generation is also performed at the word level. To form the decoder input text $x_{dec}$, we make sure all modified tokens (masked tokens and misspelled tokens) in $x_{enc}$ are replaced with [MASK] and the remaining input is kept unchanged.


The output of the encoder is the [CLS] token embedding which is taken as the first token embedding along with the $x_{dec}$ input to the decoder. The encoder pre-training loss function is the standard MLM loss, which is the token-level cross-entropy loss:
	\vspace{-4pt}
\begin{equation}\small
	\mathcal{L}_{enc} = -\frac{1}{|M|}\sum_{i\in{M}}\log P(x_i | x_{enc}, E_{\theta}),
	\vspace{-4pt}
\end{equation}
where $M$ is the set of indices of the masked tokens in $x_{enc}$. The decoder loss function $\mathcal{L}_{dec}$ is similar. The final pre-training loss is the sum of the encoder and decoder losses: $\mathcal{L}_{pt} =\mathcal{L}_{enc} + \mathcal{L}_{dec}$. 

Since the decoder has weaker language modeling power than the encoder and the number of masked tokens in $x_{dec}$ is also more than that in $x_{enc}$, the encoder needs to learn to compress as much semantic information as possible into the bottleneck [CLS] embeddings in order for this architecture to perform well in the pre-training task. We further note that there are misspelled words in $x_{enc}$ and the decoder will need to recover the original words, thus the enhanced [CLS] embeddings given by the encoder are expected to also have information about the original correctly spelled words.

Following the practice of pre-training for DRs~\cite{gao-callan-2021-condenser,lu-etal-2021-less,liu2022retromae,wang2022simlm,shen2022lexmae}, we initialize the encoder and decoder parameters with BERT and further pre-train them on the target corpus. After pre-training is finished, we disregard the decoder and only keep the encoder as the backbone encoder model for the downstream retrieval task.

	\vspace{-4pt}
\section{Multi-stage fine-tuning}
In the previous section we presented our proposed pre-training methodology for the [CLS] enhanced typo-robust backbone encoder model. Next, we introduce our fine-tuning pipeline for training an effective dense retriever. Our pipeline differs from previous methods in that it incorporates not only the commonly employed techniques of contrastive learning, hard negative mining, and knowledge distillation, but also incorporates a typos-aware fine-tuning strategy to further enhance the robustness of the DR to misspelled queries.

	\vspace{-4pt}
\subsection{Contrastive Learning for Dense  Retrieval}
Contrastive learning is a powerful technique for learning semantic representations. In the context of dense retrieval, it aims to learn representations that can differentiate between relevant and irrelevant passages for a given query. Formally, in the first stage of our fine-tuning pipeline, we fine-tune the backbone encoder model using the contrastive cross-entropy loss function as follows:
\begin{equation}\small\label{eq:ce}
	\mathcal{L}_{ce}(q)= -\log \frac{e^{\phi_{dot}(E_{\theta}(q), E_{\theta}(p^+))}}{e^{\phi_{dot}(E_{\theta}(q), E_{\theta}(p^+))} + \sum_{p^- \in P^-}e^{\phi_{dot}(E_{\theta}(q), E_{\theta}(p^-))}}
\end{equation}
where $E_{\theta}$ represents the backbone encoder model trained with our ToRoDer approach and $E_{\theta}(.)$ is the [CLS] embedding encoded by the encoder. $\phi_{dot}$ is the dot product function that computes the matching score between query [CLS] token embedding $E_{\theta}(q)$ and passage $E_{\theta}(p)$. $p^+$ is a passage judged positive (relevant) for the query $q$ and $p^-$ is a negative (irrelevant) passage. Negative passages (the set $P^-$) are sampled from the top-$k$ retrieved passages from BM25 and from in-batch negatives.

\vspace{-4pt}
\subsection{Hard Negatives Mining for Fine-tuning}
In addition to BM25 and in-batch negatives, we also use a hard negative mining approach which selects hard negative passages from the dataset using the dense retriever fine-tuned from the first stage. This approach has been widely used in previous works and has been shown to be crucial in enhancing the effectiveness of dense retrievers~\cite{xiong2020approximate,zhan2021optimizing,hofstatter2021efficiently,tonellotto2022lecture,zhao2022dense}. Therefore, we included it in the second stage of our fine-tuning pipeline.

Specifically, we employed the contrastive cross-entropy loss (Equation~\ref{eq:ce}) to train a second backbone encoder $E_{\theta}$, using hard negatives sampled from the top-k passages retrieved by the DR checkpoint trained in the first stage. This enables the optimization of the model's representations for the specific task of semantic dense retrieval, thereby resulting in an even more effective DR.

\vspace{-4pt}
\subsection{Knowledge Distillation from Cross-encoders}
Another powerful technique to improve the effectiveness of dense retrievers is Knowledge Distillation (KD) from a cross-encoder reranker. This is because the cross-encoder reranker can model the full interaction between the query and passage tokens, thus serving as a powerful teacher model for knowledge distillation~\cite{hofstatter2020improving,zhao2022dense}. We incorporate this technique in the third stage of our fine-tuning pipeline. To achieve this, we first trained a cross-encoder reranker using the Localized Contrastive Estimation (LCE) loss function~\cite{gao2021rethink}, which is similar to  Equation~\ref{eq:ce}, but where the matching scores between the query and passages are estimated by the cross-encoder reranker and hard negatives are obtained from the denser retriever trained in the second stage:

	\vspace{-6pt}
\begin{equation}\small\label{eq:lce}
	\mathcal{L}_{lce}(q)= -\log \frac{e^{Reranker(q, p)}}{e^{Reranker(q, p^+)} + \sum_{p^- \in P^-}e^{Reranker(q, p^-)}}
\end{equation}

Following this, we then applied KD training for the dense retriever by distilling the knowledge of match score distributions from the reranker teacher model using the Kullback-Leibler (KL) loss:

	\vspace{-6pt}
\begin{equation}\small
	\mathcal{L}_{kl} (q) = \tilde{S}_{reranker}(q) \cdot \log \frac{\tilde{S}_{reranker}(q)}{ \tilde{S}_{E_{\theta}}(q)}
		\vspace{-4pt}
\end{equation}
where $\tilde{S}_{reranker}$ and $\tilde{S}_{E_{\theta}}$ represent the normalized match score distributions computed by the reranker teacher model and dense retriever student model, respectively. In our implementation, we employed softmax normalization for this purpose. 
We further note that, during the KD training, (i) we use the same hard negatives used to train the reranker teacher model, (ii) the parameters of the reranker teacher model are fixed, and (iii) only the student dense retriever model is updated.

	\vspace{-4pt}
\subsection{Self-Teaching Fine-tuning}
Previous works have proposed several typos-aware fine-tuning methods, which have been shown to significantly improve the robustness of dense retrievers on misspelled queries~\cite{zhuang2021dealing,zhuang2022char,sidiropoulos2022analysing,Chen2022towards}. Thus, in addition to leveraging our typo-robust backbone encoder, in our fine-tuning pipeline we also adopt a recent typos-aware fine-tuning method called \textit{Self-Teaching} (ST)~\cite{zhuang2022char}. We do this to further improve the robustness of our model to misspelled queries.

Specifically, the ST approach is akin to the knowledge distillation method in that the training objective of ST is to minimize the KL divergence loss between the passage match score distribution obtained by the misspelled query $q'$ and the distribution obtained from the corresponding query without typos ($q$, the original query) for the dense retriever:

	\vspace{-6pt}
\begin{equation}\small
	\mathcal{L}_{ST} (q) = \tilde{S}_{E_{\theta}}(q) \cdot \log \frac{\tilde{S}_{E_{\theta}}(q)}{ \tilde{S}_{E_{\theta}}(q')}
\end{equation}
where the misspelled query $q'$ associated with each original query $q$ is obtained by executing the same text editor used in the typos-aware bottlenecked pre-training on the original queries. We explicitly incorporate the ST loss into all stages of our fine-tuning pipeline.

The final loss functions used in each stage of our fine-tuning pipeline are summarized as follows:

 \begin{itemize}
 	\item $	\mathcal{L}_{s1}(q) = \mathcal{L}_{ce}(q) + \mathcal{L}_{ST}(q) $, where negative passages are from in-batch negatives and BM25. 
 	\item $\mathcal{L}_{s2}(q) = \mathcal{L}_{ce}(q) + \mathcal{L}_{ST}(q) $, where negative passages are from in-batch negatives and the DR checkpoint trained in stage 1. 
 	\item $\mathcal{L}_{s3}(q) = \mathcal{L}_{kl}(q) + \mathcal{L}_{kl}(q')  + 
 	\tau \cdot  (\mathcal{L}_{ce}(q) + \mathcal{L}_{ST}(q)) $, where 
 	\begin{equation}\small
 		\mathcal{L}_{kl} (q') = \tilde{S}_{reranker}(q) \cdot \log \frac{\tilde{S}_{reranker}(q)}{ \tilde{S}_{E_{\theta}}(q')}
 	\end{equation}
 \end{itemize}

 In stage 3 (corresponding to loss $\mathcal{L}_{s3}(q)$), we not only distill the knowledge from the cross-encoder reranker to the dense retriever student model for the original query $q$, but also for the misspelled query $q'$. Additionally, we retain the contrastive cross-entropy loss and self-teaching loss from stages 1 and 2 to serve as regularization terms, which are controlled by the hyperparameter $\tau$.
 	
 We highlight that the loss functions detailed above are used in combination in each stage of our fine-tuning pipeline to optimize the representations of the model for the semantic dense retrieval, and to improve the robustness of the model to misspelled queries.

\section{Experimental Settings}
\subsubsection*{Datasets and Measures}
We evaluate ToRoDer using three datasets: two are publicly available, and one is internal to a large commercial web search engine and it leverages real users queries. 

For the publicly available datasets, we use the MS MARCO v1 passage dataset~\cite{bajaj2016ms} and the Typo-DL dataset~\cite{zhuang2022char}. Both datasets are based on the 8.8 million passages released with MS MARCO, and differ in the queries used. The development query set of the MS MARCO v1 passage ranking dataset contains 6,980 queries, each with an average of one relevant passage per query. To obtain misspelled queries, we used the typos generator proposed by~\citeauthor{zhuang2021dealing}~\cite{zhuang2021dealing} to randomly generate a misspelled query for each original development query. Each misspelled query is guaranteed to have exactly one typo on a non-stop word. The DL-typo dataset, on the other hand, provides 60 real misspelled queries and 60 corresponding correctly spelled queries that were corrected manually. The passage relevance judgements are provided on a 4-point scale: from 0 (Irrelevant) to 3 (Perfectly Relevant). The average number of judged passages per query is 63.52. To evaluate the effectiveness of our method and baselines, we use MRR@10 and Recall@1000 on MS MARCO dev queries and nDCG@10, MAP and Recall@1000 for DL-typo, as per common practice for these datasets.

We also use an internal dataset from a large commercial web search engine. The dataset comprises English web content, containing roughly 9.17 million passages. The dataset is divided into a training query set with 600k queries and a development query set with 31.7k queries. Each query is paired with an average of 5.96 different misspelled queries, generated from real user data. The relevance judgments are provided on a binary scale (i.e. Relevant vs Irrelevant), with an average of 1.19 relevant passages per query. It is worth noting that all misspelled queries are from real users, while their corresponding correctly spelled queries were automatically generated using our advanced internal spell-checker, and relevant passages were labeled based on users' clicks. Specifically, the dataset was compiled with the process illustrated by the following example. If there were 5 users who issued 5 different queries, and all of them clicked on the same passage, and their issued queries were corrected by the internal spell-checker to the same identical query, then we consider the corrected query as the correctly spelled version of the 5 issued queries, and the clicked passage as a relevant passage. Note that the spell-checker used to generate the potentially correct version of a query was not used in the production search pipeline.  Both the correctly spelled version of the query and the relevant passages should be considered as being a silver ground-truth because they have not undergone any extensive manual validation (though we did perform a small scale analysis to validate the correctness of a sample of queries and assessments). Edge cases can be easily constructed for which neither the corrected query corresponds to the misspelled queries, nor the relevant passage is actually relevant to the corrected query. However, in our limited manual review of the dataset, we could not identify any such cases. 
For evaluation on this dataset, we use MRR@10 and Recall@1000 on the development queries, in line with MS MARCO, because the number of relevant passages per query is similar to that dataset. More details regarding this dataset can be found in Table~\ref{table:dataset}.

To compute the effectiveness of our method and baselines, we used Ranx~\cite{DBLP:conf/ecir/Bassani22} and conducted statistical significance tests using a two-tailed paired t-test and Bonferroni correction ($p<0.01$).

\subsubsection*{Baselines}
Our selection criteria for baselines include four main factors. Firstly, we consider typos-aware DRs and the current state-of-the-art (SOTA) method in this domain. These methods are Self-Teaching~\cite{zhuang2022char} (ST) and RoDR~\cite{Chen2022towards}. Secondly, we include the SOTA DRs that use bottleneck pre-training. These methods are CoCondenser~\cite{gao2022unsupervised} and SimLM~\cite{wang2022simlm}. Thirdly, we incorporate other popular DRs that have been widely used in previous work. These methods are a BERT-based DR, ANCE~\cite{xiong2020approximate} and ColBERTv2~\cite{lin2021batch}. Lastly, we exclude DRs for which obtaining the model checkpoints is challenging, and, to ensure the reproducibility of our baselines, we rely on the publicly available model checkpoints provided by the original authors. 
This approach enables us to obtain the same results as reported in the original papers, facilitating our statistical analysis.

\begin{table}[]
		\caption{Statistics of the internal dataset from a large commercial web search engine we employ for evaluation. \vspace{-10pt}}
	\resizebox{0.55\columnwidth}{!}{
	\begin{tabular}{ll}
\hline
		\# passages & 9,167,352 \\

		\# train queries & 600,000 \\

		\# dev queries & 31,719 \\

		avg \# typo query per query & 5.96 \\

		avg \# rel doc per query & 1.19 \\

		avg query len & 4.22 \\

		avg typo query len & 4.78 \\
\hline
	\end{tabular}	
}
	\label{table:dataset}
	\vspace{-18pt}
\end{table}

\begin{table*}[!t]
	\small
		\caption{Effectiveness obtained on misspelled queries. The effectiveness of the original correctly spelled queries is presented in brackets. Methods statistically significantly better ($p < 0.01$) than others are indicated by superscripts. Models labelled with $^*$ have contributed to the DL-Typo's judgment pool (top-10 retrieved passages of these models are all judged), thus care should be put when directly comparing their effectiveness to models that have not contributed to DL-Typo, as these will have passages in the top-10 that might not be judged. \vspace{-10pt}}
	\resizebox{1\linewidth}{!}{
		\begin{tabular}{l|c|ll|lll}
			\hline
			\multicolumn{1}{c|}{\multirow{2}{*}{Models}} & \multicolumn{1}{l|}{\multirow{2}{*}{\begin{tabular}[c]{@{}l@{}}Typos-aware\\ Fine-tuning\end{tabular}}} & \multicolumn{2}{c|}{MS MARCO passage dev}                     & \multicolumn{3}{c}{DL-Typo}                                         \\ \cline{3-7} 
			\multicolumn{1}{c|}{}                        & \multicolumn{1}{l|}{}                                                                                   & \multicolumn{1}{c}{MRR@10} & \multicolumn{1}{c|}{Recall@1000} & \multicolumn{1}{c}{nDCG@10} & \multicolumn{1}{c}{MAP} & Recall@1000 \\ \hline
			
			$a$) BM25$^*$ & & 9.5 (18.7) & 61.5 (85.7) & 21.2 (52.7) & 10.4 (29.8) & 50.0 (84.4) \\
			$b$) BERT$^*$ & & 14.1$^{a}$ (32.5$^{a}$) & 69.3$^{a}$ (95.3$^{a}$) & 28.3 (72.2$^{af}$) & 16.7 (56.5$^{af}$) & 59.9 (95.7$^{af}$) \\
			$c$) BERT+ST$^*$~\cite{zhuang2022char} & \cmark & 22.6$^{abdfgh}$ (33.1$^{a}$) & 85.9$^{abdf-j}$ (94.9$^{a}$) & 43.3$^{abdhij}$ (\textbf{72.5}$^{afh}$) & 30.1$^{ab}$ (58.3$^{af}$) & 81.2$^{abd}$ (96.1$^{a}$) \\
			$d$) CharacterBERT$^*$ & & 15.6$^{ab}$ (32.7$^{a}$) & 73.0$^{ab}$ (95.0$^{a}$) & 29.7 (71.6$^{af}$) & 22.4$^{a}$ (53.8$^{a}$) & 61.8 (95.4$^{a}$) \\
			$e$) CharacterBERT+ST$^*$~\cite{zhuang2022char} & \cmark & 26.2$^{a-df-j}$ (32.5$^{a}$) & 89.7$^{a-df-j}$  (95.0$^{a}$) & 47.3$^{abdf-j}$ (70.6$^{af}$) & 34.8$^{abdf-j}$ (53.9$^{a}$) & 82.9$^{abd}$ (94.1$^{a}$) \\
			$f$) ANCE~\cite{xiong2020approximate} & & 20.3$^{abd}$ (33.0$^{a}$) & 80.2$^{abd}$ (95.9$^{acde}$) & 34.0$^{a}$ (60.6) & 24.5$^{a}$ (48.0$^{a}$) & 76.0$^{ab}$ (96.7$^{a}$) \\
			$g$) RoDR~\cite{Chen2022towards} & \cmark & 21.3$^{abdh}$ (34.3$^{a-f}$) & 84.1$^{abdfh}$ (96.1$^{a-e}$) & 35.9$^{ab}$ (70.5$^{afh}$) & 26.9$^{ab}$ (58.7$^{af}$) & 75.8$^{abd}$ (97.4$^{ae}$) \\
			$h$) TCT-ColBERTv2~\cite{lin2021batch}& & 20.3$^{abd}$ (35.8$^{a-g}$) & 80.6$^{abd}$ (96.9$^{a-g}$) & 31.1$^{a}$ (65.0$^{a}$) & 22.1$^{a}$ (56.5$^{af}$) & 74.7$^{ab}$ (98.2$^{ae}$) \\
			$i$) CoCondenser~\cite{gao2022unsupervised} & & 22.3$^{abdfgh}$ (38.1$^{a-h}$) & 84.5$^{abdfh}$ (98.4$^{a-h}$) & 30.5 (69.5$^{af}$) & 23.6$^{ab}$ (\textbf{59.2}$^{af}$) & 79.8$^{abd}$ (\textbf{99.1}$^{a-f}$) \\
			$j$) SimLM~\cite{wang2022simlm} & & 23.6$^{a-df-i}$ (\textbf{41.0}$^{a-i}$) & 83.8$^{abdfh}$ (\textbf{98.7}$^{a-h}$) & 29.9 (67.1$^{a}$) & 24.1$^{a}$ (58.1$^{af}$) & 78.7$^{abd}$ (99.0$^{abcef}$) \\ \hline
			$k$) ToRoDer & \cmark & \textbf{38.3$^{a-j}$} (40.8$^{a-i}$) & \textbf{97.8$^{a-j}$} (98.5$^{a-h}$) & \textbf{59.8}$^{a-j}$ (68.4$^{af}$) & \textbf{50.1$^{a-j}$} (58.0$^{af}$) & \textbf{95.4$^{a-j}$} (98.4$^{ae}$)\\\hline
			
		\end{tabular}
	}
\vspace{-10pt}
	\label{table:main_results}
\end{table*}

\subsubsection*{Implementation Details for ToRoDer}
For the pre-training of our ToRoDer model, we initialize the encoder $E_{\theta}$ with the  bert-base-uncased checkpoint\footnote{https://huggingface.co/bert-base-uncased} provided by Huggingface. The decoder, $D_{\theta'}$, is initialized with the last two layers of the same BERT checkpoint. We pre-train the encoder-decoder on the MS MARCO passage corpus for 80,000 steps with a learning rate of 3e-4 and a warm-up ratio of 4,000 steps. We utilized 8 NVIDIA A100 GPUs to train the model, with a batch size per GPU of 256. For the encoder-decoder ratio $\alpha$, we ensure that $30\%$ of the input tokens are used for MLM. Furthermore, for the typo injection ratio $\beta$, we set the probability of each token to be replaced with a misspelled token to $30\%$. As a result, the expected number of [MASK] tokens for the decoder is $60\%$. However, if the number of [MASK] tokens for the decoder is less than $60\%$ (for example, where there are fewer tokens replaced with misspelled tokens), we replace more tokens with [MASK] to the decoder inputs to ensure a minimum of $50\%$. We also study the impact of $\beta$ in Section~\ref{sec:beta}. Given the expensive nature of pre-training, we were unable to conduct a comprehensive grid search for the optimal values for all $\alpha-\beta$ combinations. As a result, our primary focus lies in fine-tuning the $\beta$ parameter, which is introduced for the first time in this paper. The value of $\alpha$ was already extensively examined in prior work~\cite{wang2022simlm}, and we adopt the best value reported in that study (30\%).
We use the Huggingface transformers library~\cite{wolf-etal-2020-transformers} to implement our pre-training approach. We will open-source our implementations and checkpoints upon acceptance.

In our fine-tuning pipeline, for stages 1 and 2, we train the encoder model for a total of 3 epochs with a learning rate of 2e-5 and a warm-up ratio of 1,000 steps. We set the number of negative passages for each positive passage to 15.
For stage 3, we initialize the reranker teacher model also with the \texttt{bert-base-uncased} checkpoint and train the model for 3 epochs with the same learning rate of 2e-5 and a warm-up ratio of 1,000 steps. The number of negative passages is set to 64 per positive passage. For training the student DR model, we set the hyperparameter $\tau=0.2$, learning rate to 3e-5, warmup steps to 1,000 for 4 epochs and number of negatives to 24.
We again use 8 NVIDIA A100 GPUs with a batch size of 8 per GPU to train all the models involved in the fine-tuning pipeline. We use the Tevatron DR training toolkit~\cite{Gao2022TevatronAE} for our fine-tuning pipeline and the Asyncval DR evaluation toolkit~\cite{zhuang2022asyncval} to evaluate model checkpoints generated during fine-tuning. We note that the MS MARCO training data provided by Tevatron is a title augmented version~\cite{lassance2023tale} and this is what we used in our multi-stage fine-tuning pipeline. This choice aligns with our selection of baselines, as the majority of the chosen baselines were also trained using this version of the training set. Among our 11 models, only ANCE and TCT-ColBERTv2 were trained using the original MS MARCO training data. While we acknowledge that comparing ANCE and TCT-ColBERTv2 with the other baselines may not be ideal, our primary objective is to compare against the SOTA typos-aware method CharacterBERT+ST and the SOTA DR utilizing bottleneck pre-training SimLM . These comparisons ensure a fair evaluation. 

\vspace{-10pt}
\section{Results}
\subsection{Main Results}
We begin by comparing our method with a wide range of DRs baselines, including DRs that leverage typos-aware fine-tuning methods (BERT+ST, CharacterBERT+ST and RoDR) and bottlenecked pre-training (CoCondenser and SimLM), using benchmark datasets such as MS MARCO and DL-typo. 
It is important to note that the misspelled queries in MS MARCO are artificially generated, while the misspelled queries in DL-typo are sourced from real-world query logs. The results of this comparison are reported in Table~\ref{table:main_results}.

The results indicate that all DR baselines and BM25 perform considerably worse on misspelled queries compared to correctly spelled queries; this is in line with findings in previous studies~\cite{zhuang2021dealing,zhuang2022char,sidiropoulos2022analysing,Chen2022towards}. However, fine-tuning methods that take typos into account tend to have higher effectiveness on misspelled queries, even though they may perform worse on correctly spelled queries. For example, CharacterBERT+ST has a lower MRR@10 than SimLM on MS MARCO for correctly spelled queries (32.5 vs 41.0), but it is more effective for misspelled queries (26.2 vs 23.6). This suggests that typos-aware fine-tuning strategies create a trade-off between the effectiveness on misspelled and that on correctly spelled queries. On the DL-typo dataset, this trade-off is less pronounced, and DR baselines that consider typos perform well on both types of queries. However, it should be noted that the baselines from $a$ to $e$ contributed their top-10 retrieved passages to the DL-typo judgement pool, so it may not be fair to strictly compare them directly to methods from $f$ to $k$, which did not contribute to that pool. These last methods, in fact, may retrieve, in their top-10, passages that are not judged -- and thus may be relevant (or not), but are assumed irrelevant by the evaluation, as per standard practice in IR. Despite this, among the baselines from $f$ to $j$ on DL-typo, RoDR exhibits good effectiveness on both misspelled and correctly spelled queries.

On the other hand, \textit{ToRoDer} exhibits considerable improvements in effectiveness when dealing with misspelled queries in comparison to all baselines. These improvements are consistently statistically significant. When compared to the best previous DR method that handles misspelled queries, CharacterBERT+ST, our method shows a remarkable 46\% improvement in MRR@10 and 9\% in Recall@1000 on MS MARCO. Furthermore, on the DL-typo dataset, our method outperforms the strong CharacterBERT+ST baseline across all evaluation metrics, even with an average of 2.65 unjudged passages in its top-10 retrieved passages (against no unjudged for CharacterBERT+ST). In addition, there is no other baseline method that is statistically significantly better than ToRoDer across all datasets, evaluation metrics, and query types, including when compared to SimLM on correctly spelled queries, which is the current state-of-the-art DR method on MS MARCO.

\begin{table}[t]
	\caption{Effectiveness in terms of RBP@10 with residuals on the DL-Typo dataset. Methods statistically significantly better ($p < 0.01$) than others are indicated by superscripts. Models labelled with $^*$ have contributed to the DL-Typo's judgment pool. \#  unjudged is the mean number of unjudged passages in the top-10 (third column) and the residuals provide an indication of the uncertainty in the evaluation posed by incomplete assessments. \label{table-RBP} \vspace{-10pt}}
	\resizebox{1\columnwidth}{!}{
	\begin{tabular}{l|l|p{40pt}}
\hline
		Models & RBP@10 + residual & \# unjudged  \\
		\midrule
		\textit{a)} BM25$^*$                    & 0.070 + 0.000 (0.226 + 0.000)             &          0 (0)                            \\
		\textit{b)} BERT$^*$                    & 0.125 + 0.000 (0.386$^{a}$ + 0.000)            &            0 (0)                          \\
		\textit{c)} BERT+ST$^*$~\cite{zhuang2022char}                 & 0.221$^{abhij}$ + 0.000 (0.384$^{a}$ + 0.000)       &      0 (0)                              \\
		\textit{d)} CharacterBERT$^*$           & 0.168$^{a}$ + 0.000 (0.379$^{a}$ + 0.000)           &       0 (0)                 \\
		\textit{e)} CharacterBERT+ST$^*$~\cite{zhuang2022char}      & 0.253$^{abdf-j}$ + 0.000 (0.375$^{a}$ + 0.00)    &        0 (0)                  \\
		\textit{f)} ANCE~\cite{xiong2020approximate}                    & 0.174$^{a}$ + 0.348 (0.330$^{a-eg}$ + 0.179)    &           5.58 (3.02)                           \\
		\textit{g)} RoDR~\cite{Chen2022towards}                    & 0.174$^{ab}$ + 0.217 (0.372$^{a-e}$ + 0.066)    &            3.67 (1.33)                          \\
		\textit{h)} TCT-ColBERTv2~\cite{lin2021batch}           & 0.159$^{a}$ + 0.336 (0.347$^{a-eg}$ + 0.158)    &     5.40 (2.72)                              \\
		\textit{i)} CoCondenser~\cite{gao2022unsupervised}             & 0.156$^{a}$ + 0.318 (0.373$^{a-eg}$ + 0.110)    &          5.22 (1.95)                            \\
		\textit{j)} SimLM~\cite{wang2022simlm}                   & 0.146$^{a}$ + 0.348 (0.360$^{a-eg}$ + 0.145)    &          5.63 (2.53)                            \\
		\midrule
		\textit{k)} ToRoDer                 & 0.324$^{a-j}$ + 0.156 (0.364$^{a-eg}$ + 0.116)  &     2.65 (2.00)  \\
\hline
	\end{tabular}
}
\vspace{-14pt}
\end{table}

To further analyse the uncertainty in the evaluation findings posed by the presence of biases in the completeness of assessments for the top 10 passages retrieved by the rankers on the DL-Typo dataset, we report an analysis of the residuals for Rank Bias Precision (RBP)~\cite{moffat2008rank}. We perform this residuals analysis for RBP because by design it allows their straightforward computation. Measures like nDCG and MAP, instead, are affected by issues related to normalisation when computing residuals. We instantiate RBP with the persistence parameter $\rho$ set to 0.9, meaning the user is expected to examine on average until rank position 10. We further truncate runs to rank 10, to avoid counting residual contributions beyond that rank: we thus in all effect compute RBP@10. We use the implementation of RBP provided in Ranx, which differs from the theoretical formulation of this metrics by only considering binary relevance, rather than graded relevance -- we then perform the same binarisation of relevance labels used when computing MAP.

We report RBP@10 and residuals values in Table~\ref{table-RBP}, along with statistics related to the number of unjudged passages in the top 10 results. Methods a-e have a residual of zero because all passages in the top 10 ranks are assessed. Methods affected by incomplete assessments exhibit varying magnitudes of residuals. ANCE and SimLM are among the models with highest residuals for misspelled queries, while ToRoDer exhibits the lowest residual on these queries among the models that did not contribute to the assessment pool. We highlight that a high residual does not necessarily translate to a largely higher effectiveness if assessments were complete: they rather provide a level of uncertainty in the differences in effectiveness of these methods on the DL-Typo dataset.

\begin{table}[t]
	\caption{Results obtained on our internal dataset. Zero-shot results are obtained by model checkpoints pre-trained and fine-tuned on MS MARCO (same checkpoints as those in Table~\ref{table:main_results}). ``Supervised + synthetic typo'' means checkpoints pre-trained on MS MARCO and fine-tuned on our internal dataset where misspelled queries are synthetically generated. ``Supervised + user typo'' means checkpoints pre-trained on MS MARCO and fine-tuned on our internal dataset where misspelled queries are obtained from real-world users. ``Full pipeline'' means the checkpoint is both pre-trained and fine-tuned on our internal dataset. \vspace{-10pt}}
	\centering
	\resizebox{1\linewidth}{!}{
		\begin{tabular}{l|l|ll}
			\hline
			& Methods & MRR@10      & Recall@1000 \\ \hline
			\multicolumn{1}{l|}{\multirow{4}{*}{\begin{tabular}[l]{@{}l@{}}Zero-shot\end{tabular}}}
			&$a$) BM25                       & 24.3$^{c}$ (66.3$^{bcd}$) & 63.0 (92.9$^{c}$) \\
			&$b$) CoCondenser                       & 24.0$^{c}$ (54.2$^{c}$) & 69.1$^{ac}$ (93.4$^{acd}$) \\
			&$c$) SimLM                       & 16.9 (37.4) & 64.6$^{a}$ (91.6) \\
			&$d$) ToRoDer                       & 44.5$^{abc}$ (58.5$^{bc}$) & 86.1$^{abc}$ (92.6$^{c}$) \\\hline
			\multicolumn{1}{l|}{\multirow{2}{*}{\begin{tabular}[l]{@{}l@{}}Supervised \\ + synthetic typo \end{tabular}}}
			&$e$) CoCondenser+ST            & 61.9$^{a-d}$ (83.5$^{a-d}$) & 92.0$^{a-d}$ (\textbf{97.2$^{a-df}$}) \\
			&$f$) ToRoDer (ST only)   & 74.4$^{a-eg}$ (83.6$^{a-d}$) & 94.8$^{a-eg}$ (96.9$^{a-d}$) \\ \hline
			\multicolumn{1}{l|}{\multirow{2}{*}{\begin{tabular}[l]{@{}l@{}}Supervised \\ + user typo\end{tabular}}}
			&$g$) CoCondenser+ST            & 65.6$^{a-e}$ (83.5$^{a-d}$) & 93.4$^{a-e}$ (97.1$^{a-df}$) \\
			&$h$) ToRoDer (ST only)   & 77.5$^{a-g}$ (83.8$^{a-d}$) & 95.9$^{a-g}$ (97.0$^{a-d}$) \\ \hline
			\multicolumn{1}{l|}{\multirow{2}{*}{\begin{tabular}[l]{@{}l@{}}Full pipeline \\+ synthetic typo \end{tabular}}}
			&\multirow{2}{*}{$i$) ToRoDer}   &\multirow{2}{*}{ \textbf{77.9$^{a-h}$} (\textbf{83.9$^{a-g}$})}  & \multirow{2}{*}{ \textbf{96.2$^{a-h}$} (\textbf{97.2$^{a-dfh}$})}  \\&&&\\\hline
		\end{tabular}
	}
	\vspace{-14pt}
	\label{table:bing}
\end{table}

In addition to the publicly available benchmark datasets, we also evaluate our method on our internal dataset collected from a large commercial web search engine. The results are reported in Table~\ref{table:bing}. In this study, we compare ToRoDer with BM25 and other two strong DRs, CoCondenser and SimLM. These DRs, like ToRoDer, exploit the bottlenecked pre-training approach.

We first discuss the zero-shot domain transfer results ($a$ to $d$) for the studied methods.
We find that, although DRs have much higher effectiveness than BM25 on in-domain datasets (see Table~\ref{table:main_results}), BM25 has a significantly higher MRR@10 and equal or better Recall@1000 than all the DRs on the correctly spelled queries in our internal dataset (results reported in brackets). This finding is consistent with previous studies that have highlighted the challenges faced by DRs in domain transfer tasks~\cite{chen2022out,thakur2021beir}.
Interestingly, for the misspelled queries, the effectiveness of the DRs is similar to that of BM25 except for SimLM -- SimLM performed poorly on both misspelled and correctly spelled queries in the domain transfer task. We posit that this may be due to SimLM overfitting the MS MARCO dataset; we thus omit SimLM from the reminder of the experiments in this table. Remarkably, ToRoDer exhibits a significant advantage over other  baselines on misspelled queries, even in the zero-shot setting. This suggests that ToRoDer's ability to handle misspelled queries can be transferred effectively to another domain.
\begin{table*}[]
	\caption{Comparison between ToRoDer and a traditional IR pipeline that exploits the Microsoft Bing spell-checker (MS Spellchecker). Methods statistically significantly better ($p < 0.01$) than others are indicated by superscripts. Models labelled with $^*$ have contributed to the DL-Typo's judgment pool (top-10 retrieved passages of these models are all judged), thus care should be put when directly comparing their effectiveness to models that have not contributed to DL-Typo, as these will have passages in the top-10 that might not be judged. \vspace{-10pt}}
	\resizebox{0.9\linewidth}{!}{
		\begin{tabular}{l|ll|lll}
			\hline
			\multicolumn{1}{c|}{\multirow{2}{*}{Models}}& \multicolumn{2}{c|}{MS MARCO passage dev}                     & \multicolumn{3}{c}{DL-Typo}                                         \\ \cline{2-6}
			\multicolumn{1}{c|}{}                                         & \multicolumn{1}{c}{MRR@10} & \multicolumn{1}{c|}{Recall@1000} & \multicolumn{1}{c}{nDCG@10} & \multicolumn{1}{c}{MAP} & Recall@1000 \\ \hline
			$a$) MS Spellchecker -> CharacterBERT$^*$  & 30.4$^{b}$ (32.6) & 91.6$^{b}$ (94.8) & \textbf{71.4}$^{bd}$ (\textbf{71.5}) & 53.9$^{b}$ (53.8) &95.3$^{b}$ (95.4) \\
			$b$) CharacterBERT+ST$^*$ & 26.2 (32.5) & 89.7 (95.0) & 47.3 (70.6) & 34.8 (53.9) & 82.9 (94.1) \\
			$c$) MS Spellchecker -> SimLM & \textbf{38.4}$^{ab}$ (\textbf{40.8}$^{ab}$) & 96.0$^{ab}$ (\textbf{98.5}$^{ab}$) & 65.9$^{b}$ (67.9) & \textbf{57.1}$^{b}$ (\textbf{59.0}) & \textbf{98.7}$^{b}$  (\textbf{99.0}$^{b}$) \\\hline
			$d$) ToRoDer & 38.3$^{ab}$ (\textbf{40.8}$^{ab}$) & \textbf{97.8}$^{abc}$ (\textbf{98.5}$^{ab}$) & 59.8$^{b}$ (68.4) & 50.1$^{b}$ (58.0) & 95.4$^{b}$ (98.4$^{b}$) \\\hline
		\end{tabular}
	}
	\label{table:spellchecker}
	\vspace{-10pt}
\end{table*}

We then evaluate the effectiveness of four methods for handling misspelled queries in the supervised training setting ($e$-$h$ in Table~\ref{table:bing}). Specifically, we take the CoCondenser and ToRoDer checkpoints pre-trained on MS MARCO and fine-tune them using training queries that contain misspellings. To control the experimental variables and better evaluate the impact of pre-training, we only perform the initial stage of fine-tuning, using BM25 hard negatives to train the DRs.
Methods $e$ and $f$ employ a synthetic typos-aware fine-tuning approach using synthetically generated misspelled queries, as previously reported in Table~\ref{table:main_results}. In contrast, methods $g$ and $h$ utilize misspelled queries from real-world users to perform ST training. As our results show, both CoCondenser and ToRoDer exhibit significantly higher effectiveness on correctly spelled queries after supervised fine-tuning, surpassing BM25 by a substantial margin. Additionally, MRR@10 and Recall@1000 for correctly spelled queries were found to be similar for both methods, with no statistical significance between them.
In terms of handling misspelled queries, ToRoDer shows a significantly higher effectiveness compared to CoCondenser, regardless of whether the misspelled training queries were synthetically generated or collected from real-world user query logs. Furthermore, it was observed that using real users' misspelled queries for ST training resulted in a higher effectiveness than using generated misspelled queries. Finally, we tested the full pipeline  of ToRoDer, including pre-training on our internal dataset and the three stages of fine-tuning using synthetically generated misspelled queries. The results, reported in row $i$ of Table~\ref{table:bing}, suggest that our carefully designed pre-training and fine-tuning pipeline can achieve the best effectiveness among all methods, without the need for collecting misspelled queries from real users, which often involves a costly annotation step.

\subsection{Comparison with Advanced Spell-checkers}
Next, we compare ToRoDer against the common practice of setting up a pipeline that utilizing advanced spell-checkers to correct typos in queries prior to document retrieval. For this, we consider the MS MARCO and DL-typo datasets and compare ToRoDer to the use of the Microsoft Bing spell-checker\footnote{https://learn.microsoft.com/en-us/azure/cognitive-services/bing-spell-check/overview} (MS Spellchecker) in combination with the CharacterBERT and SimLM dense retrieval models.\footnote{We did not conduct this experiments on our internal dataset since the correctly spelled queries in our internal dataset were obtained by applying a spellchecker that relies on heuristic rules to correct user queries. Therefore, using the spellchecker as a baseline would be unfair, as it was essentially used for creating the dataset itself.} The previous work from Zhuang and Zuccon performed this comparison for CharacterBERT+ST and the pipeline composed of the spell-checker and CharacterBERT~\cite{zhuang2022char}. In their analysis, they highlighted that the current state-of-the-art typos-aware fine-tuning method provided significantly lower effectiveness than the corresponding pipeline involving the production-grade spell-checker.

We report the results of this comparison in Table~\ref{table:spellchecker}. We first observe that we obtained the same results as in previous work for CharacterBERT+ST and the spell-checker. When considering ToRoDer, we observe that
our method exhibits similar effectiveness as using the spell-checker and SimLM\footnote{Note, we cannot consider a system composed of our ToRoDer without treatment for typos and a spell-checker; this is because ToRoDer naturally encapsulates the treatment of typos in the pre-training phase. We thus consider SimLM instead because it provides state-of-the-art effectiveness on correctly spelled queries.} on  MS MARCO, and outperforms the use of the spell-checker with CharacterBERT. In terms of Recall@1000, our method even exhibits statistically significant improvements over all other systems.
On the DL-typo dataset, while ToRoDer is less effective than the systems with the MS Spellchecker, the only statistically significant difference is observed for nDCG@10 when compared to the pipeline with the MS Spellchecker and CharacterBERT. It is noteworthy that the DL-typo dataset contains more judged passages for CharacterBERT than for SimLM and ToRoDer, thus these results should be interpreted with caution. On the other hand, the MS Spellchecker is trained with large scale data containing typos from real users, thus it may be more effective on DL-typo than on MS MARCO, for which misspelled queries are synthetically generated and not from real users.

Additionally, it is important to note that, from an engineering and maintenance standpoint, ToRoDer is overall a simpler system as it does not require an extra component responsible to fix typos (the spell-checker), but instead addresses them in an end-to-end manner, thus significantly simplifying the IR pipeline. This approach allows for a more streamlined and efficient process, while still achieving comparable or even superior effectiveness compared to the more popular practice of relying on a spell-checker pre-retrieval.


\subsection{Ablation Study}
We conduct several ablation experiments to study the contribution of each key component of ToRoDer's pre-training approach and multi-stage fine-tuning pipeline. Table~\ref{table:ablation} presents the results of our experiments. First, in row $a$, we remove all the components and only conduct further MLM pre-training (i.e., no encoder-decoder architecture and misspelled inputs) on the MS MARCO passage corpus for a BERT base model. We consider this as the simplest pre-training approach and thus we use it as the baseline. For row $b$, we simply add misspelled inputs to the MLM pre-training task and ask the model to recover both [MASK] and misspelled tokens (similar to the decoder pre-training loss in ToRoDer, but conducted on the encoder only architecture). Compared to row $a$, it is clear that adding the misspelled token recovery task significantly improves the model's effectiveness on misspelled queries, at the expense of marginal losses on correctly spelled queries. In row $c$, we investigate the impact of the encoder-decoder bottleneck architecture without considering the misspelled inputs. Compared to row $a$, $c$ has higher MRR@10 and Recall@1000 on correctly spelled queries but the bottleneck architecture does not help on the misspelled queries. For $d$, we use ToRoDer's pre-training checkpoint detailed in Section~\ref{sec:pretrain} and fine-tune it with BM25 hard negatives without any typos-aware fine-tuning. The results show that with our pre-training approach, the DR can already achieve higher effectiveness on misspelled queries than previous best typos-aware fine-tuning methods (compared with Table~\ref{table:main_results}), and the negative impact of the misspelled token recovery task on correctly spelled queries becomes now null (i.e., it vanishes). In order to further compare the contribution of the typo token recovery task in our ToRoDer pre-training against the advanced typos-aware fine-tuning approach, in row $e$, we remove the typo token recovery task from our pre-training and do the ST typos-aware fine-tuning. The results show that, although the ST fine-tuning brings a significant gain on misspelled queries, compared with row $d$, it is less effective than our ToRoDer pre-training. Thus, taking into account misspelled tokens at pre-training makes DRs more robust to misspelled queries than doing this at fine-tuning.
 In row $f$, which represents the DR checkpoint of our first stage fine-tuning pipeline, we observe another effectiveness boost on misspelled queries. Thus, combining  ToRoDer's pre-training with ST typos-aware fine-tuning provides significant improvements on misspelled queries, without hurting correctly spelled queries. Finally, the results in rows $g$ and $h$, which are the last two stages of our fine-tuning pipeline, suggest that the common DR fine-tuning practices of hard negative mining and knowledge distillation significantly improve the effectiveness of DRs on both misspelled and correctly spelled queries.

\begin{table}[]
	\caption{Ablation studies of the key components of ToRoDer on the MS MARCO dev set. All methods are fine-tuned with BM25 hard negatives except for $g$ and $h$, which are fine-tuned with hard negatives mined with the first stage and second stage checkpoints, respectively. Methods statistically significantly better ($p < 0.01$) than others are indicated by superscripts. Rec: Recover; Bott: Bottleneck \vspace{-10pt}}
	\label{table:ablation}
	\resizebox{1\linewidth}{!}{
		\begin{tabular}{l|>{\centering}p{0.4cm}>{\centering}p{0.4cm}>{\centering}p{0.3cm}>{\centering}p{0.3cm}>{\centering}p{0.3cm}|ll}
			\hline
			Note & Rec. & Bott. & ST & HN & KD & MRR@10      & Recall@1000 \\\hline
			$a$) MLM pre-training&&&&&& 19.9 (37.0) & 81.0 (98.0) \\
			$b$) &\cmark &&&&& 31.7$^{ace}$ (36.5) & 95.1$^{ace}$ (97.7) \\
			$c$) &&\cmark&&&& 19.8 (37.8$^{ab}$) & 80.8 (98.2$^{b}$) \\
			$d$) ToRoDer pre-training&\cmark&\cmark&&&& 33.0$^{abce}$ (37.7$^{ab}$) & 96.0$^{abce}$(98.1$^{b}$) \\
			$e$)&&\cmark&\cmark&&& 27.4$^{ac}$ (37.4) & 91.4$^{ac}$ (97.9)\\
			$f$) First stage fine-tuning&\cmark&\cmark&\cmark&&& 35.1$^{a-e}$ (37.7$^{b}$) & 97.3$^{a-e}$ (98.0) \\
			$g$) Second stage fine-tuning&\cmark&\cmark&\cmark&\cmark&& 36.9$^{a-f}$ (38.9$^{a-f}$) & 97.6$^{a-e}$ (98.4$^{bef}$) \\\hline
			$h$) ToRoDer full pipeline&\cmark&\cmark&\cmark&\cmark&\cmark& \textbf{38.3}$^{a-g}$ (\textbf{40.8}$^{a-g}$) & \textbf{97.8}$^{a-f}$ (\textbf{98.5}$^{abef}$) \\\hline
		\end{tabular}
	}
\vspace{-10pt}
\end{table}

\subsection{Impact of the Amount of Typos in Queries}\label{sec:beta}

Next, we analyse the impact of the amount of typos present in individual queries during pre-training and inference on the effectiveness of the DR model.
For this, we setup experiments on the MS MARCO dev queries and control the amount of misspelled words in each query.
For the pre-training task, we control this through the typo injection ratio parameter $\beta$.
We vary $\beta$ from 0 (no typos -- equivalent to row $c$ in Table~\ref{table:ablation}) to 1.0 (every input token is misspelled and no [MASK] tokens are present in the inputs), with a step size of 0.1. This results in 11 pre-training checkpoints for ToRoDer, which we fine-tune using the MS MARCO training data without any typos-aware fine-tuning approach.

\begin{figure}
	\centering
		\includegraphics[width=0.49\columnwidth]{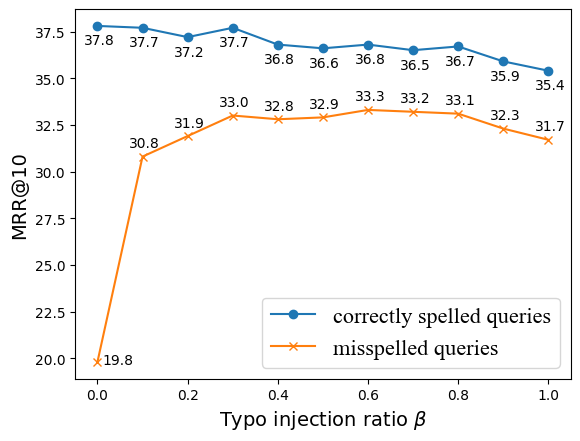}	
		\includegraphics[width=0.49\columnwidth]{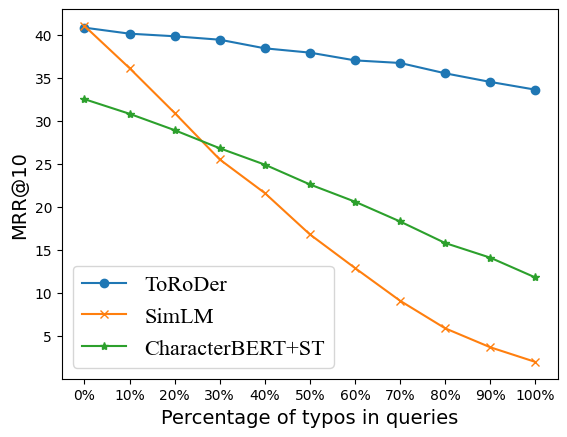}
\vspace{-10pt}
\caption{(Left:) Impact of typo injection ratios $\beta$. Fine-tuning performed with BM25 hard negatives (no ST typos-aware fine-tuning). (Right:) Effectiveness across percentages of typos. 0\%: queries without typos (original queries). 100\%: every word is misspelled. \label{fig:2}}
\vspace{-10pt}
\end{figure}

Figure~\ref{fig:2} (Left) illustrates the effect of the typo injection ratio $\beta$ used in ToRoDer's pre-training on the downstream retrieval.
Injecting a small number of typos into the input text, e.g., $\beta=0.1$, significantly increases the effectiveness of the model on misspelled queries, increasing MRR@10 from 19.8 to 30.8. However, the effectiveness gain on misspelled queries plateaus after $\beta > 0.3$ and gradually decreases as $\beta$ increases thereafter. On the other hand, the effectiveness on correctly spelled queries starts to decrease after $\beta>0.3$. Therefore, in previous experiments we set $\beta=0.3$ for ToRoDer's pre-training as it strikes a good balance between effectiveness on misspelled and on correctly spelled queries.

In Figure~\ref{fig:2} (Right), we evaluate the inference effectiveness of the fully trained ToRoDer in response to different amounts of typos in queries. ToRoDer's effectiveness on MS MARCO dev queries is compared to that of CharacterBERT+ST, which employs typos-aware fine-tuning, and SimLM, which does not use any typos-aware pre-training or fine-tuning.
We observe that the effectiveness of SimLM decreases dramatically as the amount of typos increases: SimLM's effectiveness becomes extremely poor when every word in the queries is misspelled. Although CharacterBERT+ST has a lower MRR@10 than SimLM when there are only few typos in the queries, it is better than SimLM when 30\% or more of the query words are misspelled. On the other hand, ToRoDer  exhibits strong robustness across all amounts of misspellings. The slope of the curve is clearly less than that of CharacterBERT+ST and ToRoDer still performs well even when every single query word is misspelled (100\%). This suggests that ToRoDer is more robust to different amounts of typos in queries than CharacterBERT+ST and SimLM.

\section{Conclusion and future works}
We presented ToRoDer, a novel pre-training approach for dense retrievers that improves their robustness against misspelled queries by incorporating a typos-aware bottlenecked mechanism into the pre-training process.
ToRoDer's pre-training approach uses the encoder-decoder bottlenecked architecture and the typo token recovery pre-training task to enhance the [CLS] token representations of the BERT model, making it not only encapsulate richer semantic information but also making it more robust to misspelled text inputs. Our experimental results on publicly available benchmark datasets and an internal commercial web search engine dataset indicate that ToRoDer, when combined with advanced multi-stage dense retriever fine-tuning and typos-aware fine-tuning methods, can achieve a remarkable level of effectiveness on misspelled queries. Furthermore, ToRoDer achieves comparable effectiveness to retrieval pipelines that incorporate a complex commercial spell-checker component, while achieving this in an end-to-end manner.

We believe future work should consider modifying the backbone PLM model used in ToRoDer to rely on CharacterBERT~\cite{el2020characterbert} rather than BERT, as in previous work by~\citeauthor{zhuang2022char}~\cite{zhuang2022char}. They in fact illustrated that the BERT tokenizer creates large data distribution shifts when handling misspelled text inputs compared to when handling correctly spelled text, and that CharacterBERT provides a more suitable backbone model for this task. We believe that by selecting the appropriate backbone PLM, pre-training approach, and fine-tuning methods, further improvements can be made to address the low robustness of dense retrievers on misspelled queries. We also plan to investigate the effectiveness of ToRoDer when applied to other PLM backbones, like RoBERTa, ELECTRA and the likes, and evaluate how ToDoDer performs with respect to other query variations rather than just misspelled queries. 

We have made our code and trained models publicly available at \url{https://github.com/ielab/boder}.


\balance
\bibliographystyle{ACM-Reference-Format}
\bibliography{sigir2023}

\appendix

\end{document}